\documentclass[journal,transmag]{IEEEtran}

\usepackage[pdftex]{graphicx}
\graphicspath{{../pdf/}{../jpeg/}}
\DeclareGraphicsExtensions{.pdf,.jpeg,.png}
\usepackage{amsmath}
\usepackage{array}
\usepackage{amssymb}
\usepackage[pdftex,colorlinks=true,bookmarks=false,citecolor=blue,urlcolor=blue]{hyperref}
\usepackage{caption}
\usepackage{subcaption}
\usepackage{multicol}
\begin{document}

\title{Fabrication and Characterization of a Mode-selective 45-Mode Spatial Multiplexer based on Multi-Plane Light Conversion}

\author{\IEEEauthorblockN{Satyanarayana Bade\IEEEauthorrefmark{1}, Bertrand Denolle\IEEEauthorrefmark{1}, Gauthier Trunet\IEEEauthorrefmark{1}, David Allioux\IEEEauthorrefmark{1},\\ Pu Jian\IEEEauthorrefmark{1}, Olivier Pinel\IEEEauthorrefmark{1} and Guillaume Labroille\IEEEauthorrefmark{1}} \IEEEauthorblockA{\IEEEauthorrefmark{1}CAILabs\\}
		38 boulevard Albert 1er, 35200 Rennes, France	
	\thanks{Corresponding author: D. Allioux (email: david@cailabs.com).}}


\IEEEtitleabstractindextext{
\begin{abstract}
Space Division Multiplexing (SMD) is a very attractive technique for addressing the ever-growing demands in transmission capacity by enabling the use of a new parameter \textemdash\ space \textemdash\ to increase the number of channels in multi-mode fibers. One key component to build a spatially multiplexed-based optical network is a spatial multiplexer and demultiplexer combining signals from multiple single-mode fibers into as many channels in a multi-mode fiber. In this article, we report the fabrication and characterization of a pair of 45-mode spatial multiplexer and demultiplexer saturating all the modes of a standard 50~$\mu$m core graded-index (OM2) multi-mode fiber. The multiplexers are based on Multi-Plane Light Conversion (MPLC), a technique that enables the control of the transverse shape of the light by multiple reflections on specifically designed phase plates. We show that by using a separable variable basis of modes, such as Hermite-Gaussian (HG) modes, we are able to drastically reduce the number of reflections hence reducing the insertion losses and modal crosstalks. The multiplexers typically show an average 4~dB insertion loss and -28~dB cross-talk across the C band. Finally, we emphasize the use of this higher-order modes multiplexer to explore the propagation properties inside multi-mode fibers and more specifically the mode group crosstalks as well as the impact of fiber bending.
\end{abstract}

}
\maketitle
\IEEEdisplaynontitleabstractindextext
\IEEEpeerreviewmaketitle

\section{Introduction}

\IEEEPARstart{I}{nformation}-based technologies are driving a durable demand in terms of transmission capacity meaning that the limits of current optical networks could shortly be reached~\cite{Richardson2013}. One promising approach to increase the capacity of optical fibers is the use the spatial diversity like Mode Division Multiplexing (MDM) in few-mode or multi-mode fibers (MMFs)~\cite{Richardson2013,Ryf2015,Soma2017}. Several multiplexing methods have already been proposed among which mode selective photonic lanterns \cite{AmezcuaCorrea2016,LeonSaval2014}. However, though photonic lanterns up to 15 modes have been demonstrated experimentally, the selectivity rapidly decreases with the number of modes due to the complexity of fiber engineering~\cite{Velazquez2018}.

Multi-Plane Light Conversion (MPLC) is a mode-selective multiplexing technique that has already proven highly efficient for 6, 10 and 15 channels \cite{Labroille2014,Labroille2016,Barre2017}. This approach is based on free-space conversion between orthonormal input and output bases of spatial modes through spatial phase transforms generated by a succession of phase plates. Previous work have shown promising results over the full C+L band exhibiting low insertion losses (IL) of 4.4~dB,  and very low modal crosstalks (XT) of $-23$~dB~\cite{Labroille2014,Labroille2016,Barre2017}. One major benefit of this technology is its high versatility as it is capable to address a large variety of modes from Laguerre-Gaussian (LG), Linearly-Polarized (LP) or Hermite-Gaussian (HG) modes to orbital angular momentum (OAM) \cite{Saad2017} modes with both fiber or free space output; it is also compatible with high-power beam shaping processes \cite{Garcia2017}. However, previous works have shown that the more modes are multiplexed, the more phase transforms are required with the MPLC technology, thus complicating the assembly of the multiplexer and decreasing its optical performances. Recently, the use of a triangular spot arrangement in input and of a Hermite-Gaussian modal basis in output has been reported to drastically decrease the number of phase transforms \cite{Fontaine2017}. This new approach led to a leap in terms of mode numbers with the demonstration of a pair of multiplexer and demultiplexer with up to 45 modes \cite{Bade2018} and the generation of 210 modes with a spatial light modulator \cite{Fontaine2018}, proving that the number of modes is not a challenge anymore. In this article, we extend the results already presented in \cite{Bade2018}. We show mathematically that the number of phase transforms can be drastically reduced for a special type of spatial modes, namely a separable basis of modes. We then demonstrate experimentally how MPLC with separable bases in input and output enables the design of multiplexers selectively addressing modes of graded-index fibers with a reduced number of phase transforms. We report the fabrication and characterization of a pair of 45-mode multiplexer and demultiplexer addressing all the 45 modes of a 50~$\mu$m core graded-index multi-mode fiber with an average of 4~dB insertion loss and -28~dB of crosstalk across the full C-band. By saturating all the modes guided by the fiber, the multiplexers are finally used to study mode group propagation inside the multi-mode fiber. This enables us to investigate the modal crosstalk between all mode groups as well as visualize the impact of fiber bending on the modal cross-talk within a conventional multi-mode fiber.

\section{Multi-Plane Light Conversion with separable mode basis}

Multi-Plane Light Conversion (MPLC) is a technique that allows performing any unitary spatial transform. Theoretically, MPLC enables the lossless conversion of any set of $N$ orthogonal spatial modes into any other set of $N$ orthogonal modes through a succession of transverse phase profiles separated by free space propagation serving as a fractional Fourier transform operation. For a given conversion, the phase profiles can be calculated by a wavefront matching algorithm~\cite{brevet}.

\begin{figure}[ht]
	\centering
	\begin{subfigure}[b]{0.45\textwidth}
		\centering
		\includegraphics[width=1\textwidth]{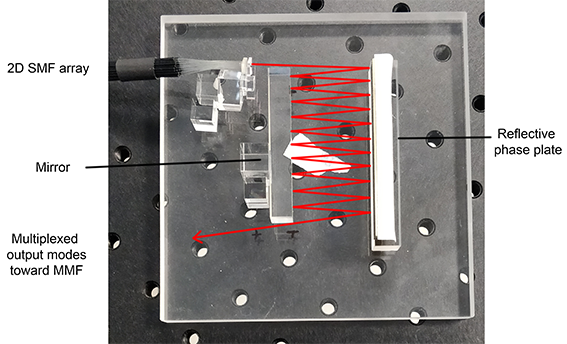}
	\end{subfigure}
\caption{\label{fig:mplc}Implementation of a 45-mode MUX based on MPLC: Picture of a 45-mode MUX based on MPLC of separable variable modes. The beam path is shown in red.}
\end{figure}

In particular, MPLC enables mode selective spatial multiplexing, i.e. the conversion of $N$ spatially separate input Gaussian beams into $N$ orthogonal propagation modes of a few-mode or multi-mode fiber~\cite{Labroille2014}. Practically, MPLC is implemented using a multi-pass cavity, in which the successive phase profiles are all manufactured on a single reflective phase plate (see Fig.~\ref{fig:mplc}). The cavity is formed by a mirror and the reflective phase plate and performs the successive phase profiles. An MPLC used in the reverse direction implements the demultiplexing operation.

The number of phase profiles required for a given MPLC is a trade-off between the number of modes, the complexity of the phase and amplitude profile of the modes, the ease of assembly of the multi-pass cavity, and the performance of the MPLC. In particular, the IL increases with the number of phase profiles due to coating losses and phase error on the phase plate as well as optical component imperfections.
We have previously demonstrated an increase in the number of phase profiles required for $N$-mode multiplexers with constant performance when $N$ increases: for average IL $<$5~dB and average XT $<-21$~dB, the number of phase profiles required for 6, 10 and 15 mode multiplexers were respectively 7, 14 and 20~\cite{Labroille2014,Labroille2016,Barre2017}.

In a previous work, we have theoretically and experimentally demonstrated that the number of phase profiles can be strongly reduced for a special class of input and output modes bases \cite{Bade2018}, namely separable variable mode bases. We here extend this approach and develop the concept of separable mode bases introduced in the aforementioned article. Let us consider an electromagnetic field $E(x,y,z)$  propagating in the z-direction of a Cartesian space $(xyz)$.  The field is said to be with separable variables if $E$, at a given plane $z=0$, can be written as:
\begin{equation*}
E(x,y,z=0) = f_{z=0}(x) g_{z=0}(y).
\end{equation*}

One can show that in the Fresnel approximation, the field can be written at any point in time and space as \cite{Goodman}: 
\begin{equation*}
E(x,y,z) = \frac{e^{ikz}}{i\lambda z}\iint^\infty_{-\infty}E(x',y',0)e^{\frac{ik}{2z}[(x-x')^2+(y-y')^2]}dx'dy'
\end{equation*}
meaning that the field remains with separable variables along the propagation in $z$ and can be written as 
\begin{equation*}
E(x,y,z) = f_z(x)g_z(y)
\end{equation*}
with 
\begin{align*}
	f_z(x)& =\sqrt{\frac{e^{ikz}}{i\lambda z}}\iint^\infty_{\infty}f_0(x')e^{\frac{ik}{2z}(x-x')^2}dx'\\
	g_z(y)& =\sqrt{\frac{e^{ikz}}{i\lambda z}}\iint^\infty_{\infty}g_0(y')e^{\frac{ik}{2z}(y-y')^2}dy'
\end{align*}
where $f_z$ and $g_z$ are independent functions completely defined by the value of the field in $z=0$, $f_0(x')$ and $g_0(y')$~\cite{Goodman}.

Let us now introduce a spatial phase change with a separable phase profile $\Phi(x,y)= \psi(x) + \theta(y)$. The output field is the product $E(x,y,z)\times e^{i\Phi(x,y)} = (f_z(x)e^{i\psi(x)})(g_z(y)e^{i\theta(y)})$. The phase modification along the $x$ (resp. $y$) direction only impacts the horizontal (resp. vertical) component. This means that, after the phase conversion, the resulting field remains with separable variables, justifying the separable phase profile denomination. We can now introduce the concept of separable basis as a basis of dimension $N^2$ with vectors defined as the products of two terms with separable variables $\{E_{n,m}(x,y)=f_n(x)g_m(y)\}_{1\leq n,m\leq N}$. We want to emphasize the fact that the two functions forming the separable product are themselves indexed by $n$ and $m$. 

MPLC converts the field from an $N$ orthonormal input bases to an $N$ orthonormal output one. By using the aforementioned properties of separable phase change, the conversion of a separable basis of $N^2$ modes $\{E_{n,m}(x,y)=f_n(x)g_m(y)\}_{1\leq n,m\leq N}$ to a separable basis of output modes $\{E'_{n,m}(x,y)=f'_n(x)g'_m(y)\}_{1\leq n,m\leq N}$ can be achieved with separable phase profiles, hence reducing the complexity of the MPLC to that of an $N$-mode problem. Indeed, if we convert each input mode $E_{n,m}$ to its equivalent  output $E'_{n,m}$, one can find a collection of $M$ 1D phase profiles $\{\psi^{(k)}(x)\}_{1\leq k\leq M}$ enabling the MPLC of $\{f_n(x)\}$ into $\{f'_n(x)\}$, and similarly a collection $\{\theta^{(k)}(y)\}_{1\leq k\leq M}$ enabling the MPLC of $\{g_m(y)\}$ into $\{g'_m(y)\}$. The collection of $M$ 2D phase profiles $\{\Phi^{(k)}(x,y)=\psi^{(k)}(x)+\theta^{(k)}(y)\}_{1\leq k\leq M}$ enables the MPLC of $\{E_{n,m}(x,y)\}_{1\leq n,m\leq N}$ into $\{E'_{n,m}(x,y)\}_{1\leq n,m\leq N}$.
With this approach, the MPLC of a separable basis of $N^2$ modes requires approximately the same number of phase profiles as the multiplexing of $N$ arbitrary modes. This greatly reduces the implementation complexity of the multiplexer without compromising the performance.

\section{Design and fabrication of a 45-mode multiplexer with separable mode basis}

The guided modes of a multi-mode fiber with a parabolic refractive index profile, such as a standard 50~$\mu$m core graded-index multi-mode fiber (so-called 50/125~$\mu$m fiber or OM2 fiber) can be analytically written as Hermite-Gaussian modes $HG_{nm}(x,y,z)$~\cite{Choudary1977,Beijersbergen1993}. One can easily show that these HG modes form a separable mode basis. 

\begin{figure}[ht]
	\centering
	\begin{subfigure}[b]{0.45\textwidth}
		\centering
		\includegraphics[width=\textwidth]{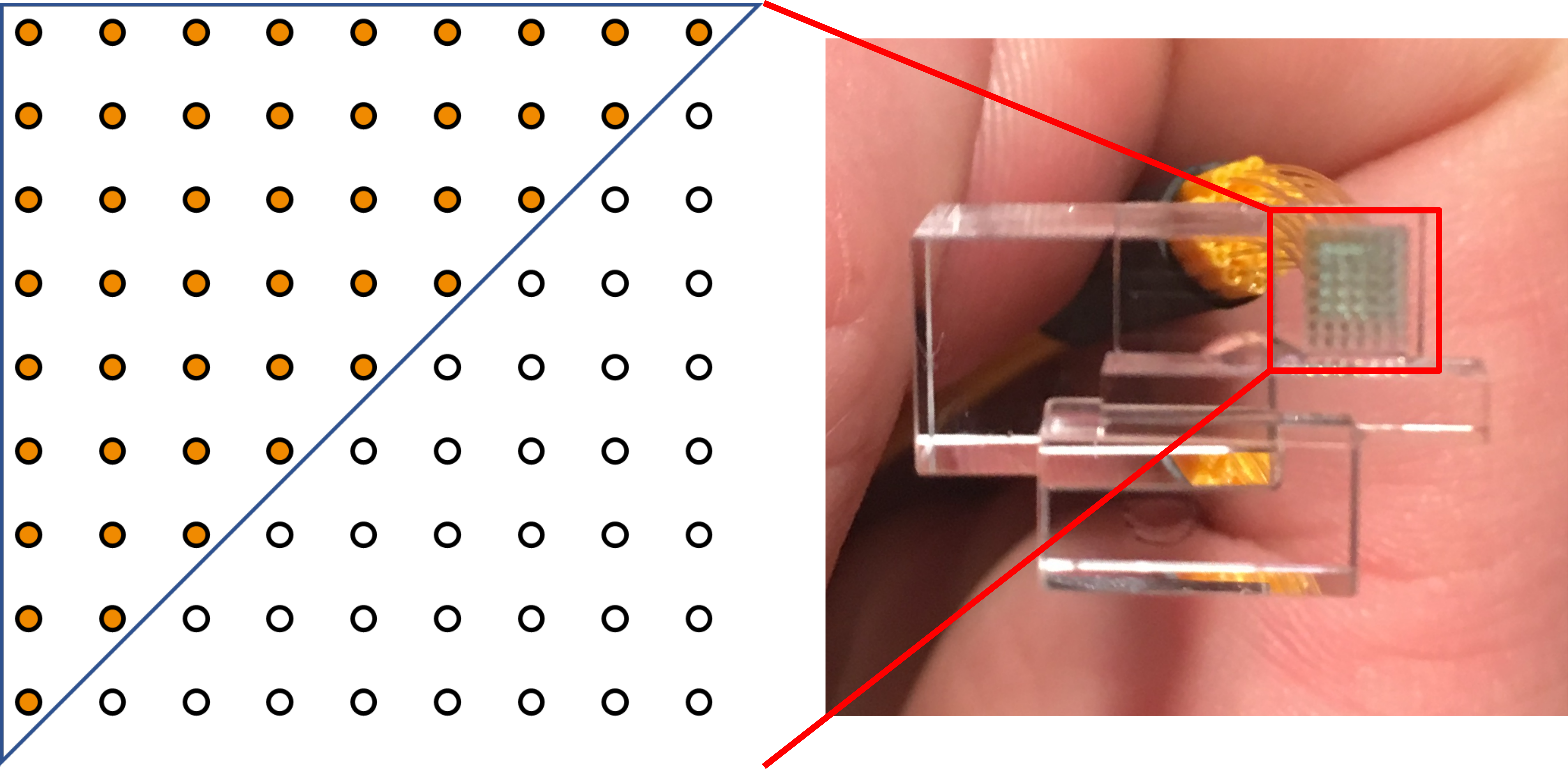}
	\end{subfigure}
	\caption{\label{fig:array_DA}Left, triangular configuration inscribed inside the square arrangement. Right, 45-fiber array.}
\end{figure}

To build a separable input basis, we arranged a 2D array of single mode fibers (SMFs) in a $9\times 9$ square matrix configuration. Among them, we have chosen 45 SMFs arranged in a triangle inscribed in the square as shown in Fig.~\ref{fig:array_DA}. Given that all the input SMFs are identical and that the grid of fibers has a fixed pitch, it is easy to show that the triangle configuration of input modes forms a spatially separable variable basis.

It follows that the MPLC transforming a 2D array of SMFs to all the 45 modes of a 50/125~$\mu$m MMF requires approximately the same number of phase profiles as multiplexing 9 arbitrary spatial modes. The 45 modes of the fiber corresponds to the 9 first mode groups with $0\leq n+m<9$ constant inside a mode group. As was shown in~\cite{Fontaine2017}, this justifies the choice of a triangular configuration of SMFs as an input and of HG modes basis as an output to drastically reduce the number of phase profiles reflection by using separable variable properties.

\begin{figure}[ht]
	\begin{subfigure}[b]{0.5\textwidth}
		\centering
		\includegraphics[width=0.79\textwidth]{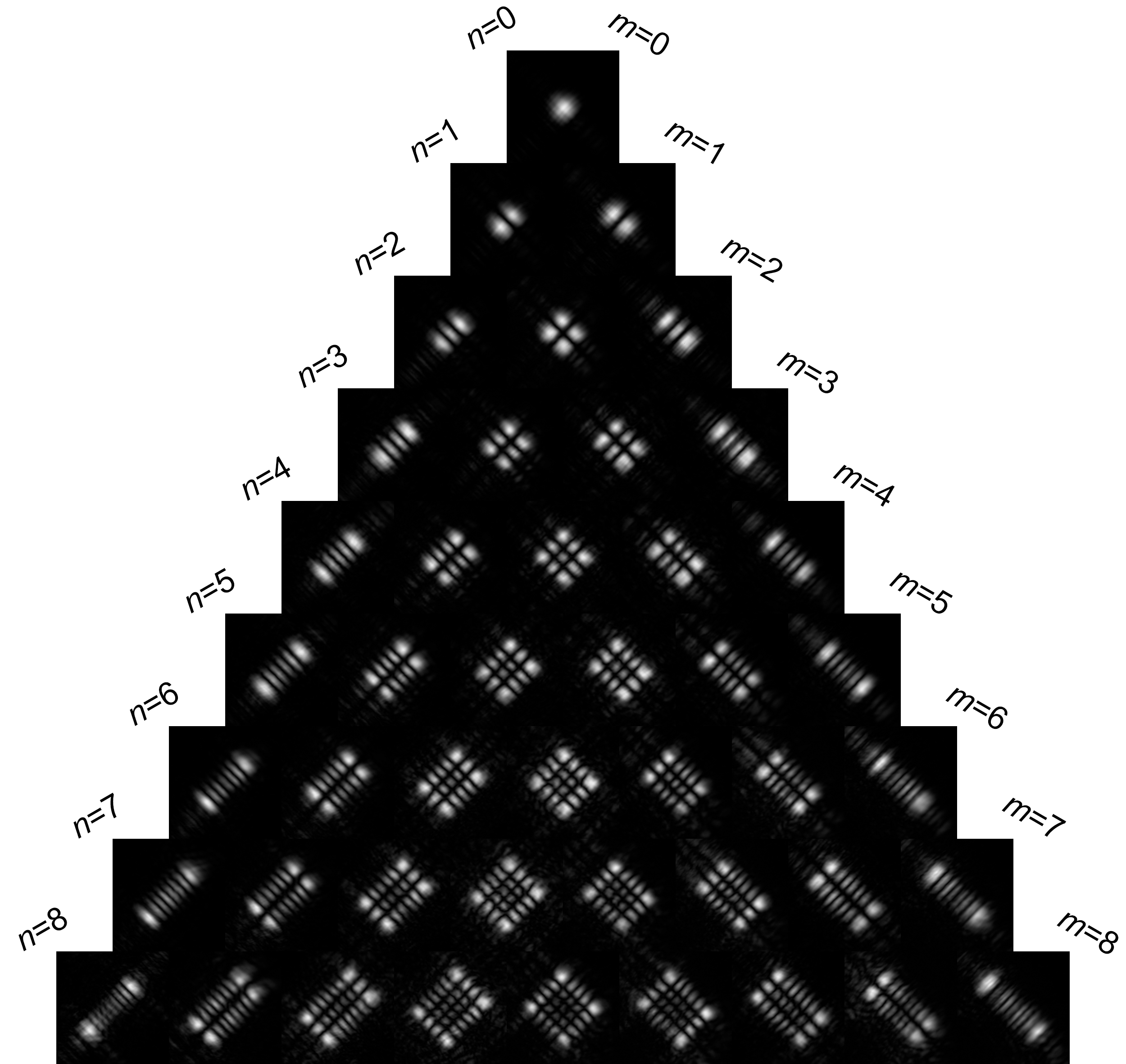}
	\end{subfigure}
	\caption{\label{fig:modes}Experimentally measured free-space output $HG_{nm}$ modes of the 45-mode multiplexer.}
\end{figure}

To demonstrate this, we fabricated a pair of 45-mode multiplexer and demultiplexer, both identical. On the input side the triangular array serves as the separable bases. The multi-pass cavity consists of 11 phase profiles inscribed on the reflective phase plate. On the output side, the light is coupled to the 50/125~$\mu$m graded index fiber. The spatial mode profiles in free-space, before coupling into the multi-mode fiber, are shown in Fig.~\ref{fig:modes}. The images were obtained by exciting the modes with a light source consisting of a superluminescent diode centered at 1550~nm with FWHM of 50~nm and were captured by a near-infrared camera with Indium Gallium Arsenide sensor.

\section{Performance of the 45-mode multiplexer and demultiplexer}

\begin{figure}[ht]
	\centering
	\begin{subfigure}[b]{0.45\textwidth}
		\centering
		\includegraphics[width=\textwidth]{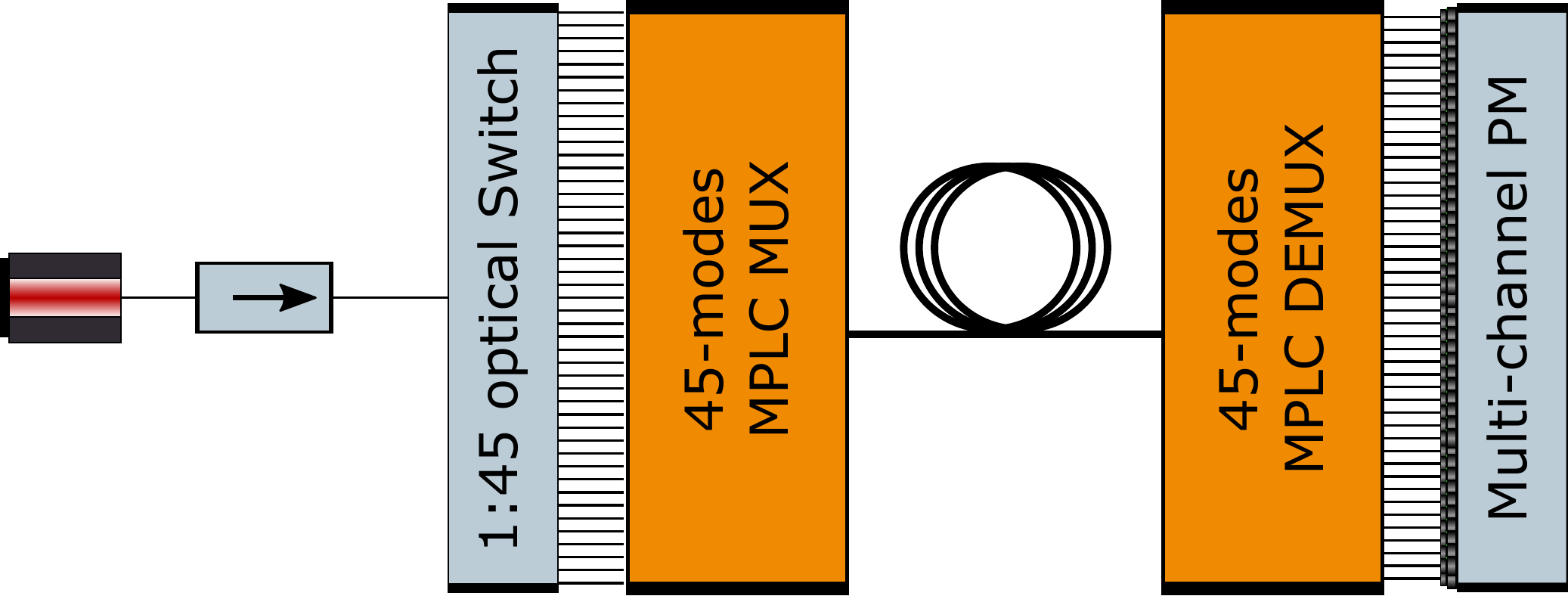}
	\end{subfigure}
	\caption{\label{fig:Charac}Characterization setup: The light source is a tunable laser (DFB). After an optical isolator, an optical switch and a multi-channel optical power meter (PM) are used to characterized the back-to-back (MUX + MMF + DEMUX) system.}
\end{figure}

\begin{figure}[ht]
	\centering
	\begin{subfigure}[b]{0.40\textwidth}
		\centering
		\includegraphics[width=\textwidth]{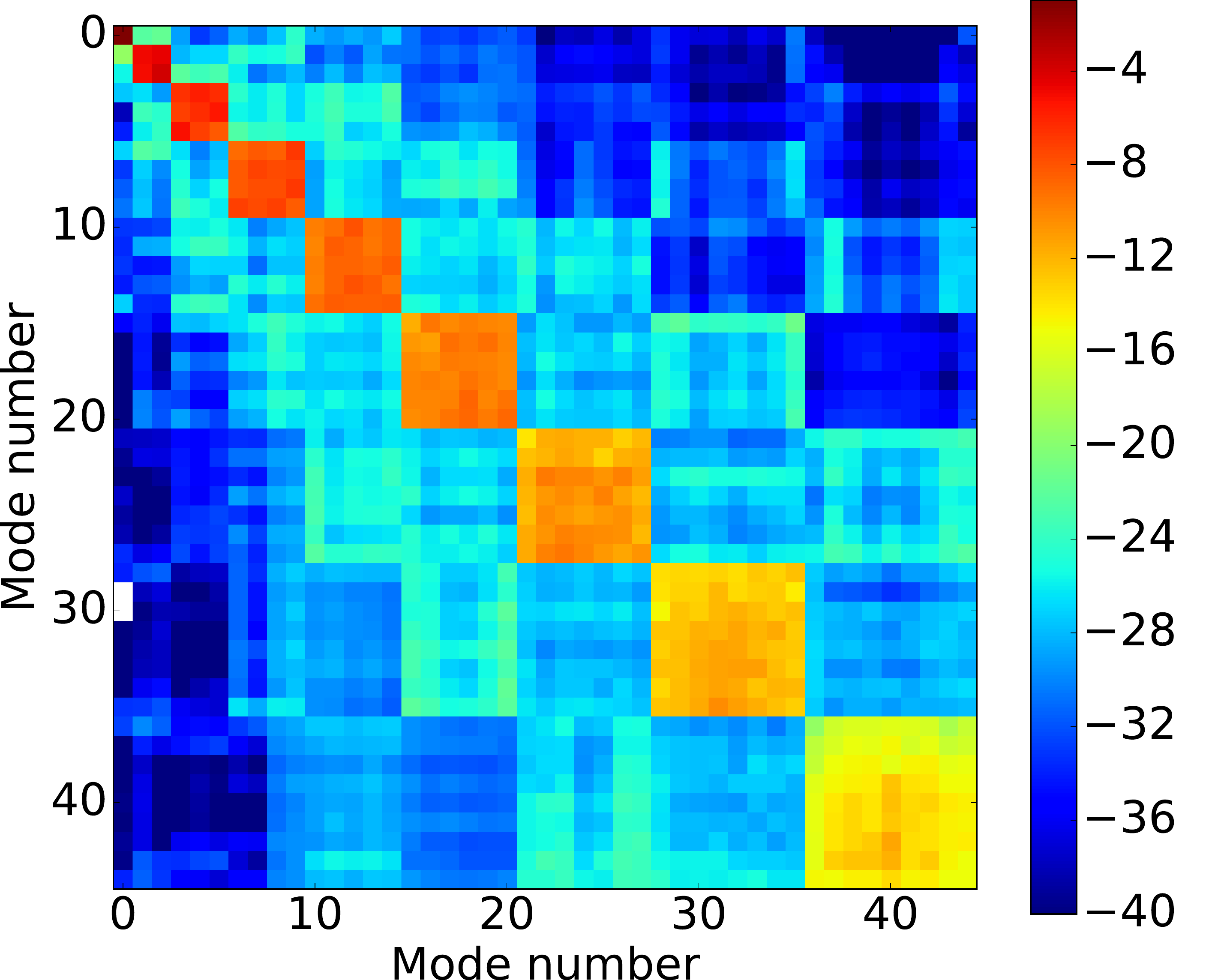}
	\end{subfigure}
	\caption{\label{fig:rawmatrix} Output power transmission matrix in dBm of the back-to-back system comprising the 45-mode MUX, 20~meters of OM2 fiber and the 45-mode DEMUX.}
\end{figure}

The 45-mode pair system is characterized by measuring the transmission matrix of a back-to-back system comprising a multiplexer, 20 meters of MMF and a demultiplexer. The measurement setup is similar to the one described in~\cite{Labroille2016}. We used a tunable distributed feedback (DFB) source covering the entire C+L band (1530 nm to 1630 nm) as an input. A 1:45 optical switch enables to sequentially excite the 45 input channels of the MUX. For each input mode, we measure the output power of all 45 output channels after the DEMUX using a multi-channel power-meter as shown in figure \ref{fig:Charac}. The resulting transmission matrix is shown in figure \ref{fig:rawmatrix}. 

Inside GI fibers, $HG_{n,m}$ modes with constant $n+m$ indices - forming a row in figure ~\ref{fig:modes} - present degenerated propagation constants. Degenerated modes have a very strong interaction and couple easily to each other, forming a mode group numbered $(n+m+1)$. On the contrary, modes from different groups have low interactions. This intra-mode group-mixing is illustrated by the block diagonal behavior of the transmission matrix in figure \ref{fig:rawmatrix}.

In order to neglect this intra-group mixing effect from the fiber in the measurement of the performances of the multiplexer and demultiplexer, we average the insertion loss and the modal cross-talk inside a mode-group. We consider the same input power $P_{in}$ for each input, and we write $P_{(n,m)\to(p,q)}$ the output power of the DEMUX channel of mode $HG_{p,q}$ when the MUX channel of mode $HG_{n,m}$ is lit. For an input mode $HG_{n,m}$, we calculate the coupling efficiency as the sum of the output powers of modes inside the same group $n+m$ and divide it by $P_{in}$. As such, the coupling efficiency can be written as:

\begin{equation}
\eta_{n,m} = \frac{\sum_{p+q=n+m}P_{(n,m)\to(p,q)}}{P_{in}}
\end{equation}

As an example for the input mode $HG_{01}$: 

\begin{equation}
\eta_{0,1} = \frac{P_{(0,1)\to(0,1)}+P_{(0,1)\to(1,0)}}{P_{in}}
\end{equation}

Similarly, the crosstalk is averaged between all outputs of the same mode groups. The crosstalk between the input mode $HG_{n,m}$ and any output mode of mode group $(k+l+1)$ is written as: 

\begin{equation}
XT_{HG_{n,m}\rightarrow HG_{k,l}} = \frac{\sum_{u+v = k+l}P_{(n,m)\to(u,v)}}{(k+l+1)\times \sum_{p+q = n+m}P_{(n,m)\to(p,q)}}.
\end{equation}

As an example, the crosstalk between input $HG_{0,0}$ and the modes of mode group 2 can be written as:

\begin{eqnarray}
XT_{HG_{0,0}\rightarrow HG_{0,1}} & = & XT_{HG_{0,0}\rightarrow HG_{1,0}} \\
 & = & \frac{P_{(0,0)\to(0,1)}+P_{(0,0)\to(1,0)}}{2\times P_{(0,0)\to(0,0)}}
\end{eqnarray}

\begin{figure}[ht]
	\begin{subfigure}[b]{0.49\textwidth}
		\centering
		\includegraphics[width=\textwidth]{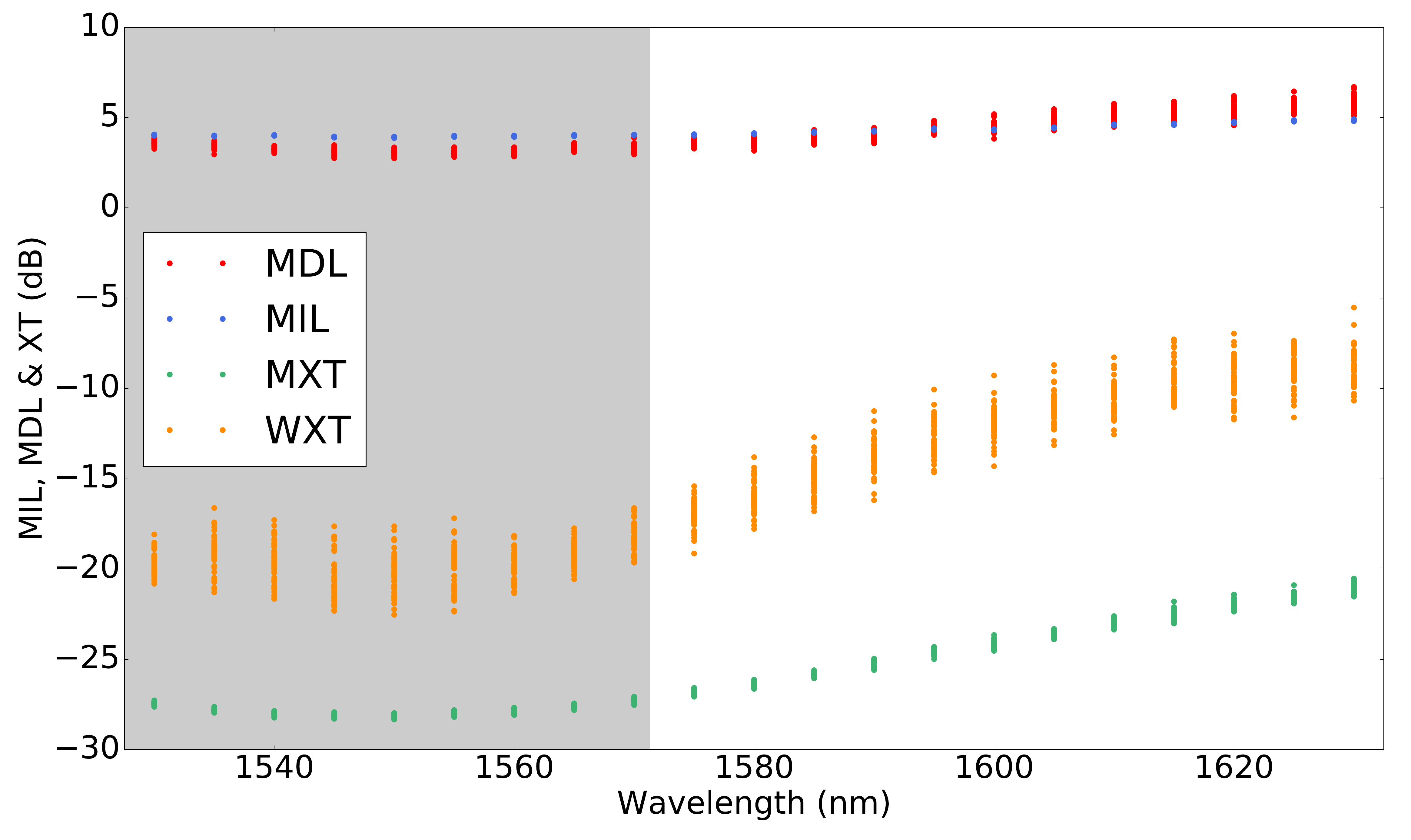}
	\end{subfigure}
\caption{\label{fig:dfb_results} IL (in blue), MDL (in red), average XT (in green) and maximum XT (in orange) of a single MUX for all measured transmission matrices.}
\end{figure}

Assuming that the multiplexer and demultiplexer are identical, the IL per MUX at 1550~nm ranges from 3.1~dB to 6.1~dB with an average of 4~dB and the average mode-to-mode XT is -28~dB. We then record the performances across the full C+L band (1530 to 1630~nm). For each wavelength, we measure 50 transmission matrices while bending and twisting the MMF in order to test various multi-path interference conditions. The IL, MDL, and XT for these 50 measurements at each wavelength are shown in Fig.~\ref{fig:dfb_results}. Over the C band (1530 to 1570~nm), the average insertion loss is 4~dB, the average XT is -28~dB, and the maximum MDL and XT observed over all the 450 measurements are respectively 3.6~dB and -18~dB.

\section{Exploring fiber properties with high order modes multiplexer}

The development of a spatially multiplexed network typically based on MDM or Mode Group Division Multiplexing (MGDM) \cite{Franz2012,Chen2011,Ryf2014} requires not only spatial multiplexers but also a precise knowledge of the propagation modes inside the fiber. In a realistic network, fiber bending and twisting also affect the propagation inside the multi-mode fiber. However, the influence of propagation-induced and bending-induced mode dependent loss and modal cross-talk in multi-mode fibers has so far been little documented. A few articles recently reported the influence of bending losses in few-mode fibers but are limited to the study of the first 6 LP modes due to the limits of the multiplexing technology used \cite{Huang20165,Leandro2015,Zheng2016,Xu2017}. In this last section, we show that the 45-mode highly selective multiplexer enables us to investigate the propagation properties of all the modes inside standard 50~$\mu$m graded index fibers. To illustrate this, we study the transmission matrix after 2~km of propagation inside a standard OM2 fiber to examine the modal content and the mode interaction. We also use the multiplexers to investigate the impact of bending on mode dependent loss and modal XT and evaluate the sensitivity to external perturbations. 


\begin{figure}[ht]
	\centering
	\begin{subfigure}[]{0.49\textwidth}
		\centering
		\includegraphics[width=\textwidth]{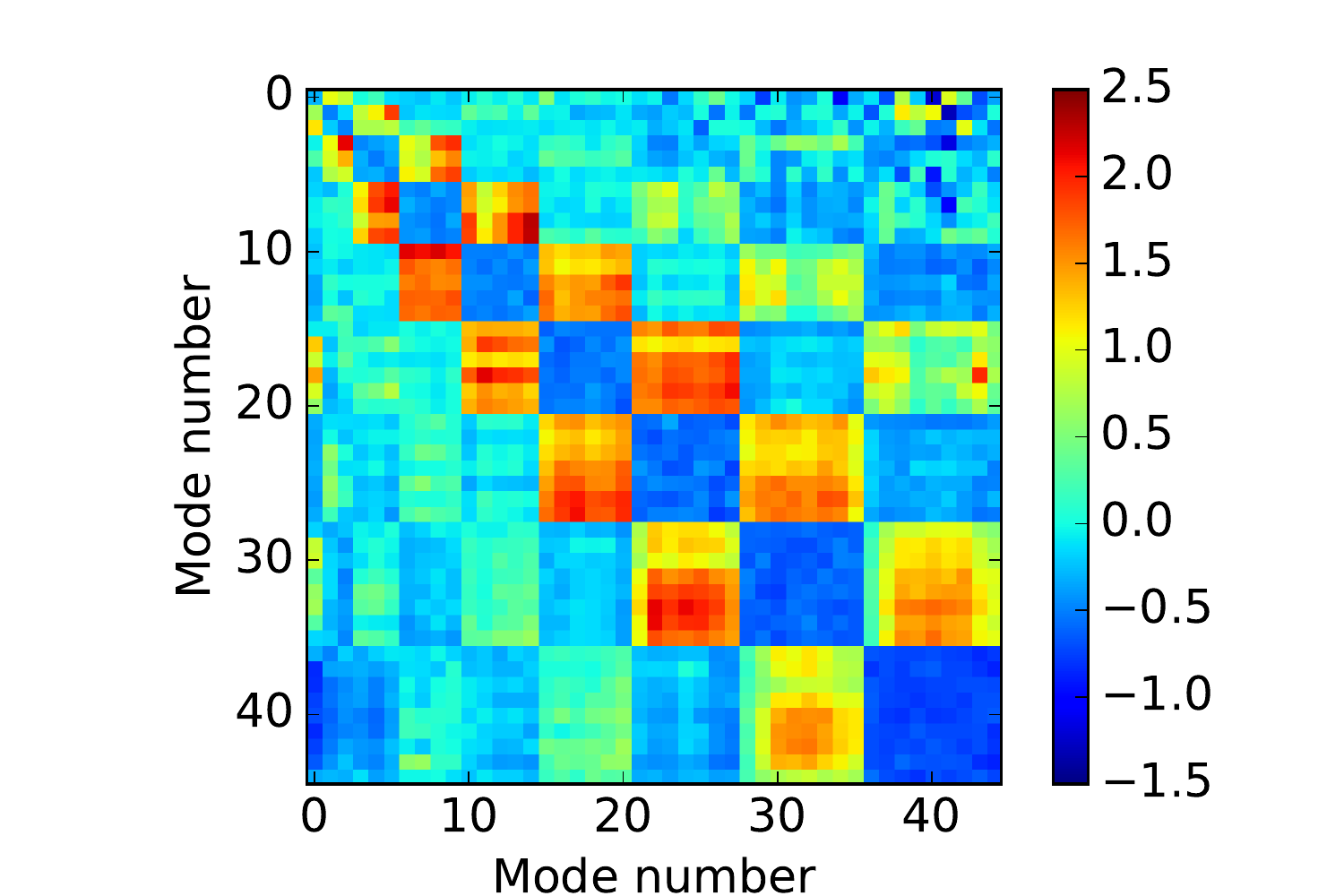}
	\end{subfigure}
	\caption{\label{fig:losses_2km_vs_1m_dB} Output matrix power difference for 2~km and 1~m propagation distance in dB.}
\end{figure}

We begin with studying the propagation-induced XT between mode groups. As mentioned previously, the modes inside a single group are degenerate and exchange energy with each other. On the contrary, the crosstalk with modes from other groups is expected to be low. To investigate experimentally propagation-induced loss and cross-talk, we compare the transmission power matrices after 1~m and 2~km of propagation in a standard OM2 fiber. After 1~m propagation, modes inside the same mode group mix with each other, but the crosstalk with other groups is very low. This measurement can serve as a reference since the impact of crosstalk between mode groups is mainly due to the MPLC. We calculate the output power matrix after 2~km and compare it to the reference matrix. The result is presented in Fig.~\ref{fig:losses_2km_vs_1m_dB}: the blue colors correspond to power loss while orange-red corresponds to power gain. From this result, we can extract a linear inter mode-group cross-talk crosstalk per kilometer for each mode group and report an average -0.35~dB/km maximum mode-group loss measured for the highest order (9th) mode group. We observe that the energy is predominantly exchanged with the nearest mode group as can be seen by the high modal gain in the nearest diagonal groups. 

\begin{figure}[ht]
	\centering
	\begin{subfigure}[]{0.24\textwidth}
		\centering
		\includegraphics[trim = 1.7cm 0.5cm 1.2cm 0cm, clip, width=\textwidth]{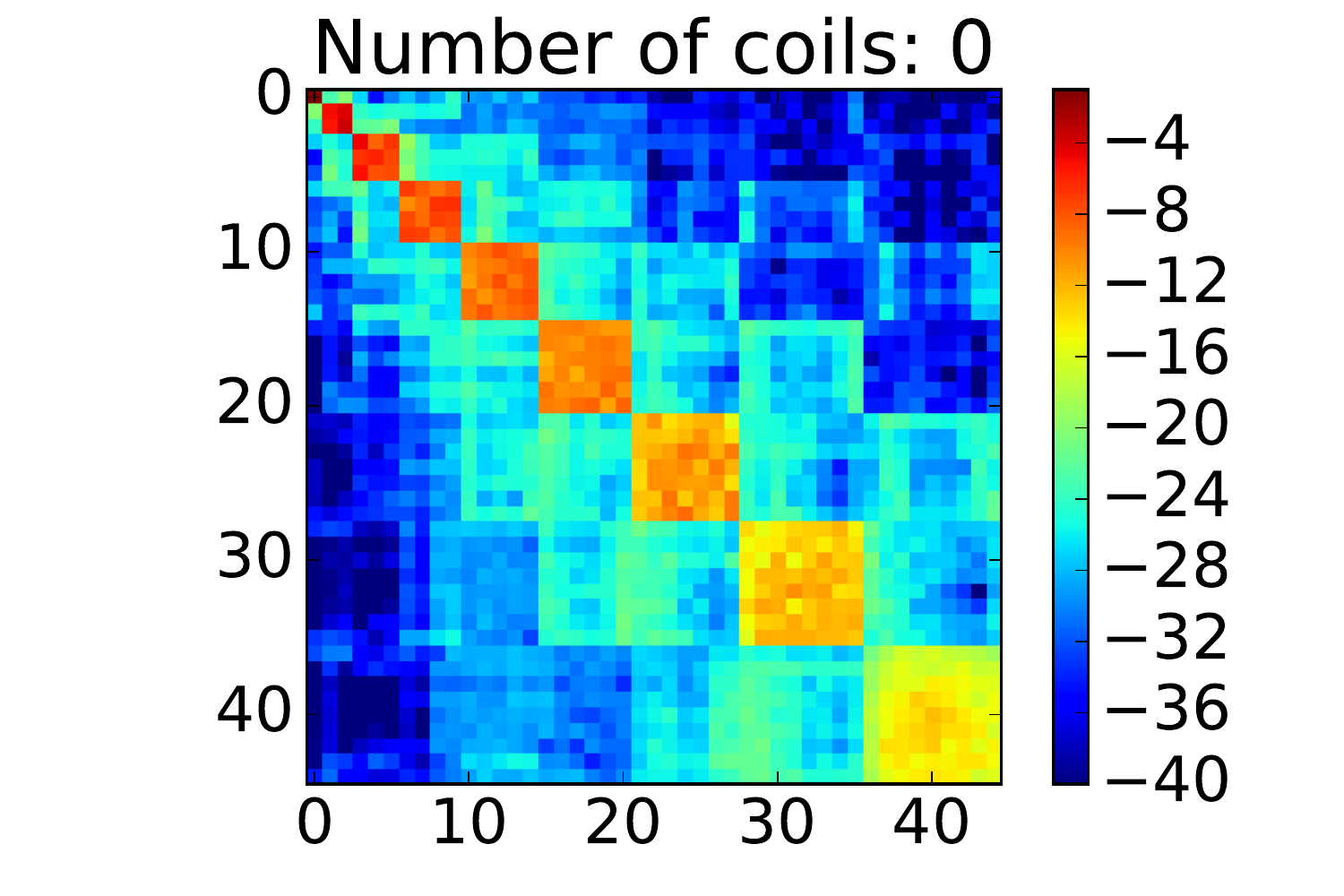}
		\caption{}\
	\end{subfigure}
	\begin{subfigure}[]{0.24\textwidth}
		\centering
		\includegraphics[trim = 1.7cm 0.5cm 1.2cm 0cm, clip, width=\textwidth]{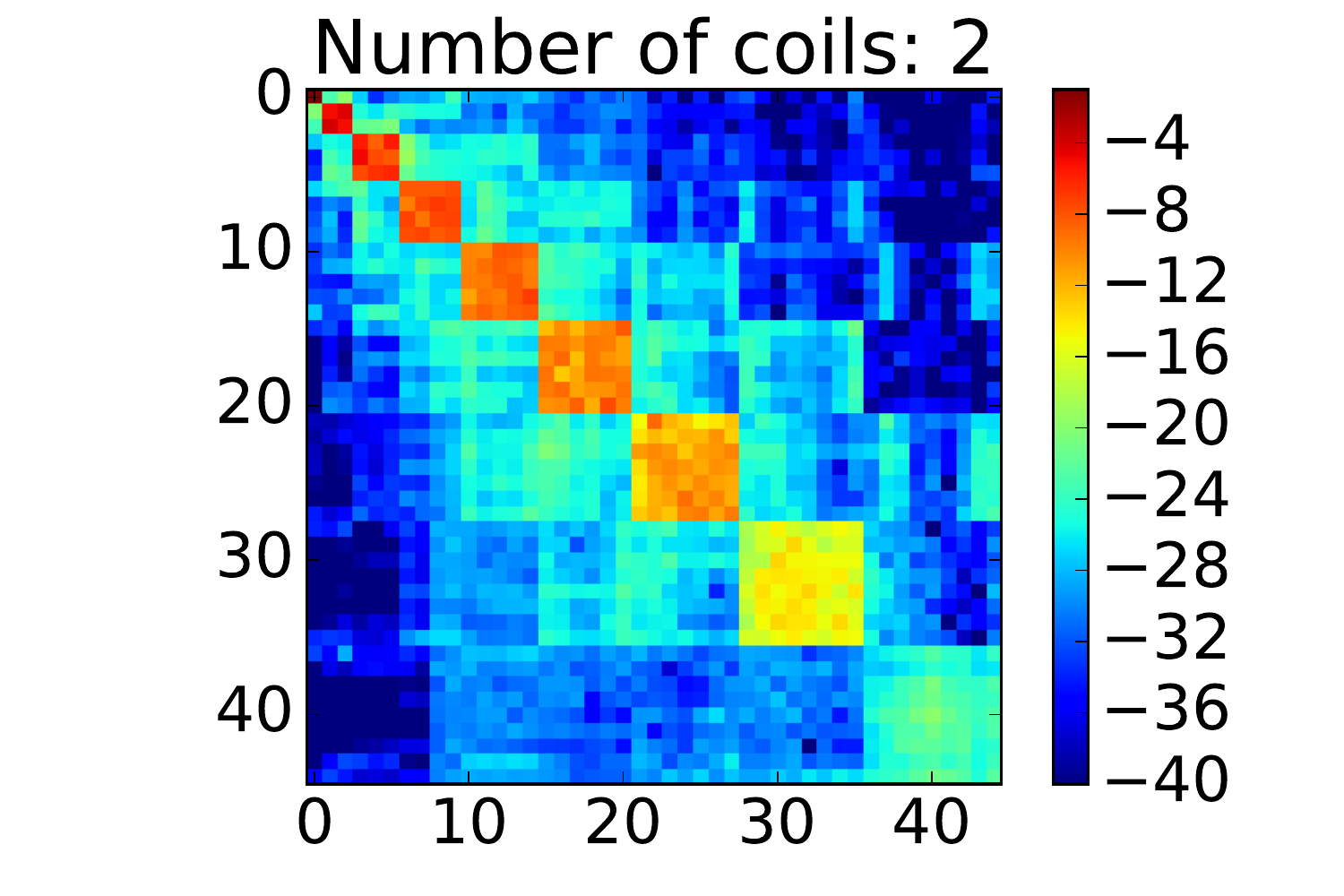}
		\caption{}\
	\end{subfigure}\\
	\begin{subfigure}[]{0.24\textwidth}
		\centering
		\includegraphics[trim = 1.7cm 0.5cm 1.2cm 0cm, clip, width=\textwidth]{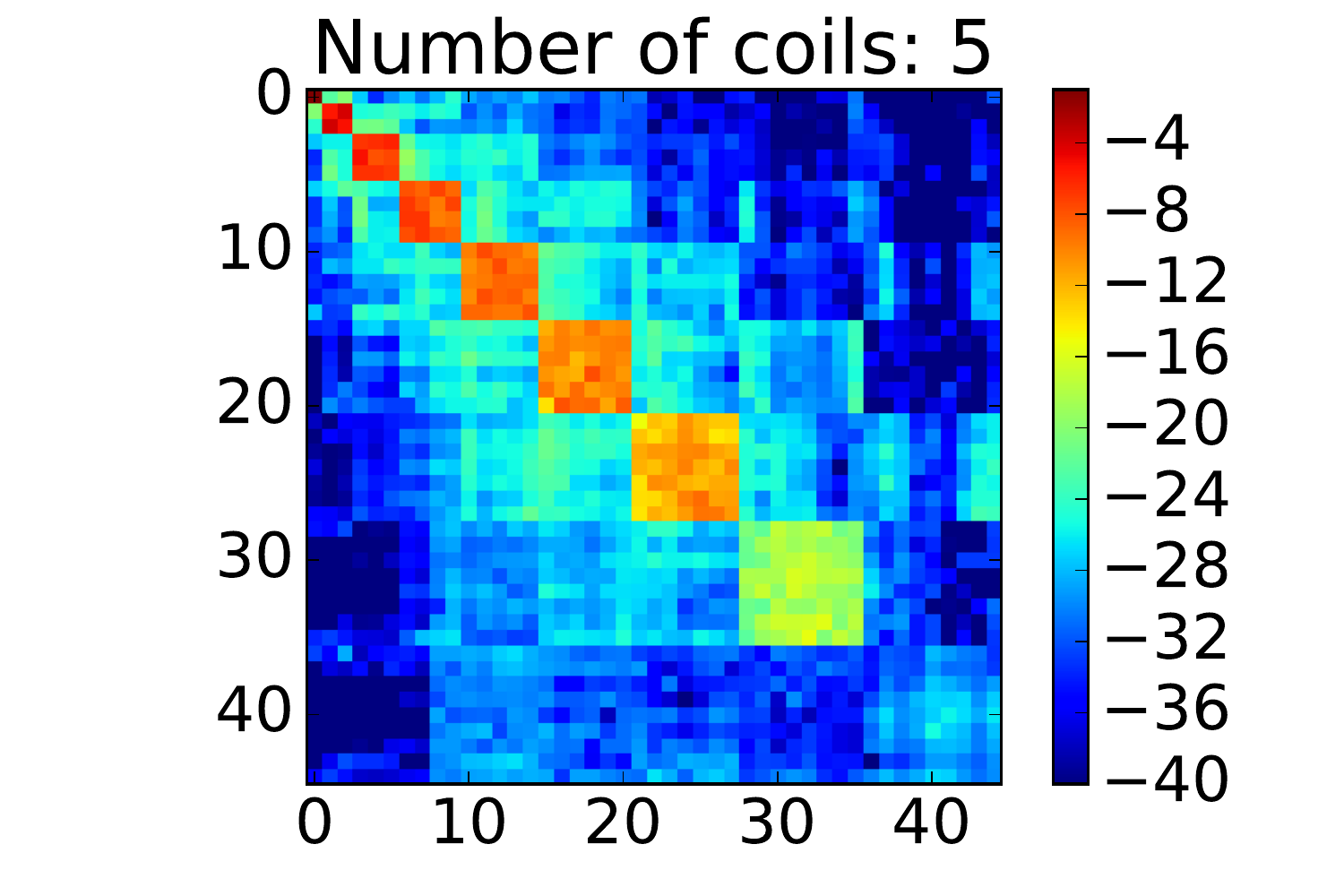}
		\caption{}\
	\end{subfigure}
	\begin{subfigure}[]{0.24\textwidth}
		\centering
		\includegraphics[trim = 1.7cm 0.5cm 1.2cm 0cm, clip, width=\textwidth]{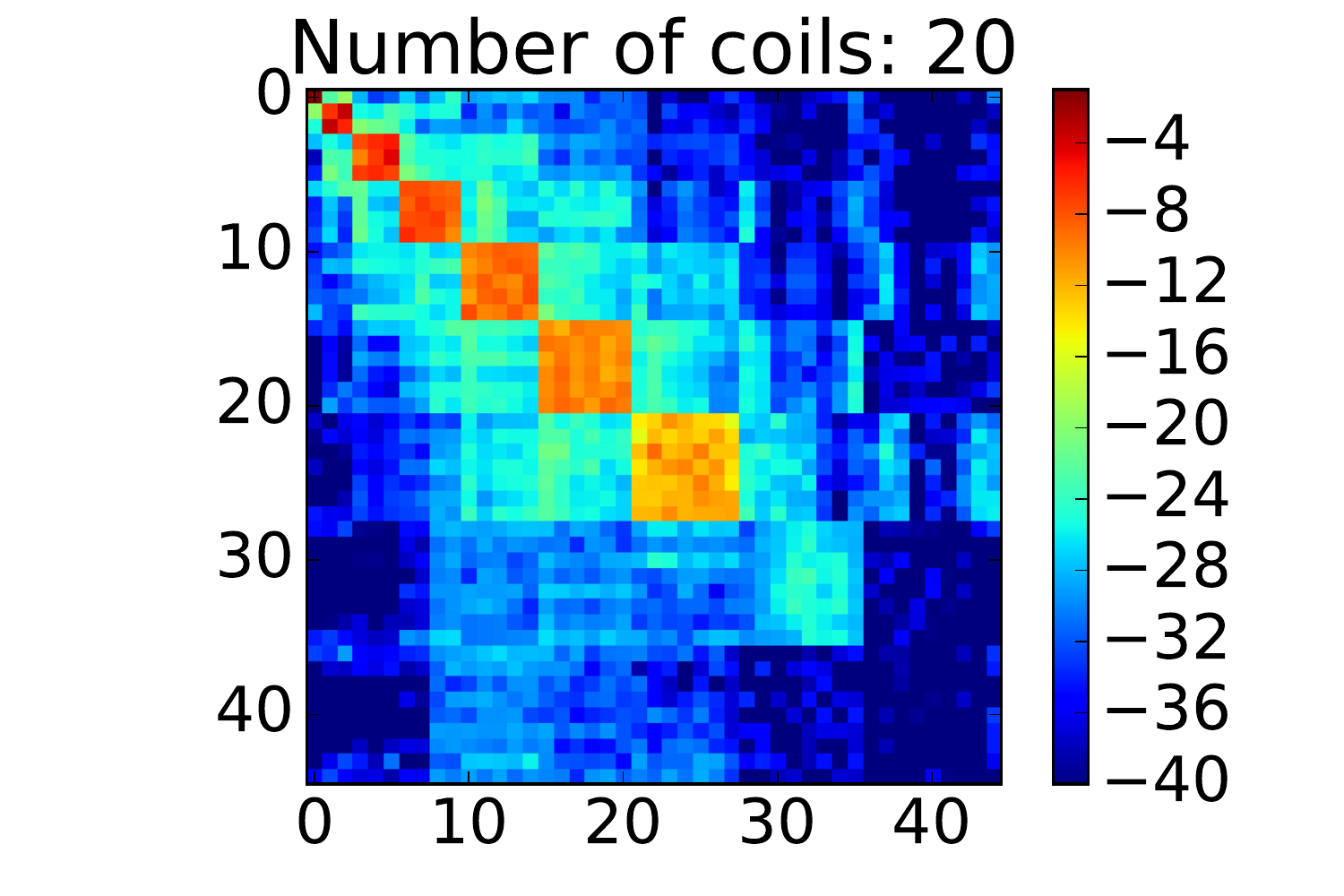}
		\caption{}\
	\end{subfigure}
	\caption{\label{fig:coil} Output power matrices in dBm of the 45 HG-modes propagating in an OM2 fiber coiled around a 12.7~mm diameter rod for (a) 0, (b) 2 (c) 5 and (d) 20 numbers of coil.}
\end{figure}

We then study the impact of bending. We record the output power matrix while coiling the fiber around rod of diameter 12.7~mm and 5~mm diameter. As the fiber is coiled tighter, losses induced by scattering increases. Higher order modes, having larger effective area and lower effective refractive index are less confined and thus experience more losses. This behavior can be assimilated to a high order mode filter. We conduct the first experiment on the 12.7~mm diameter rod. The results are shown in Fig.~\ref{fig:coil} with the number of coils increasing from 0, meaning that the fiber is straight, to 20 coils. These results show that the output power of each mode group decreases with a larger number of coils. The loss in dBm is expected to be linear \cite{koyamada1998} with a different slope for each mode, stronger for higher order modes. When loss is too high, the output power reaches the noise floor as it is the case for the 9th mode group with approximately 5 coils. Typically for the same number of coils, the propagation loss of the 8th mode group is impacted and the transmission vanishes for 20 coils. On the contrary, the lower order mode groups are preserved and the impact on them is almost negligible, even for a large number of coils, indicating a very high confinement of these modes.\

\begin{figure}[ht]
	\centering
	\begin{subfigure}[]{0.24\textwidth}
		\centering		
		\includegraphics[trim = 1.7cm 0.5cm 1.2cm 0cm, clip, width=\textwidth]{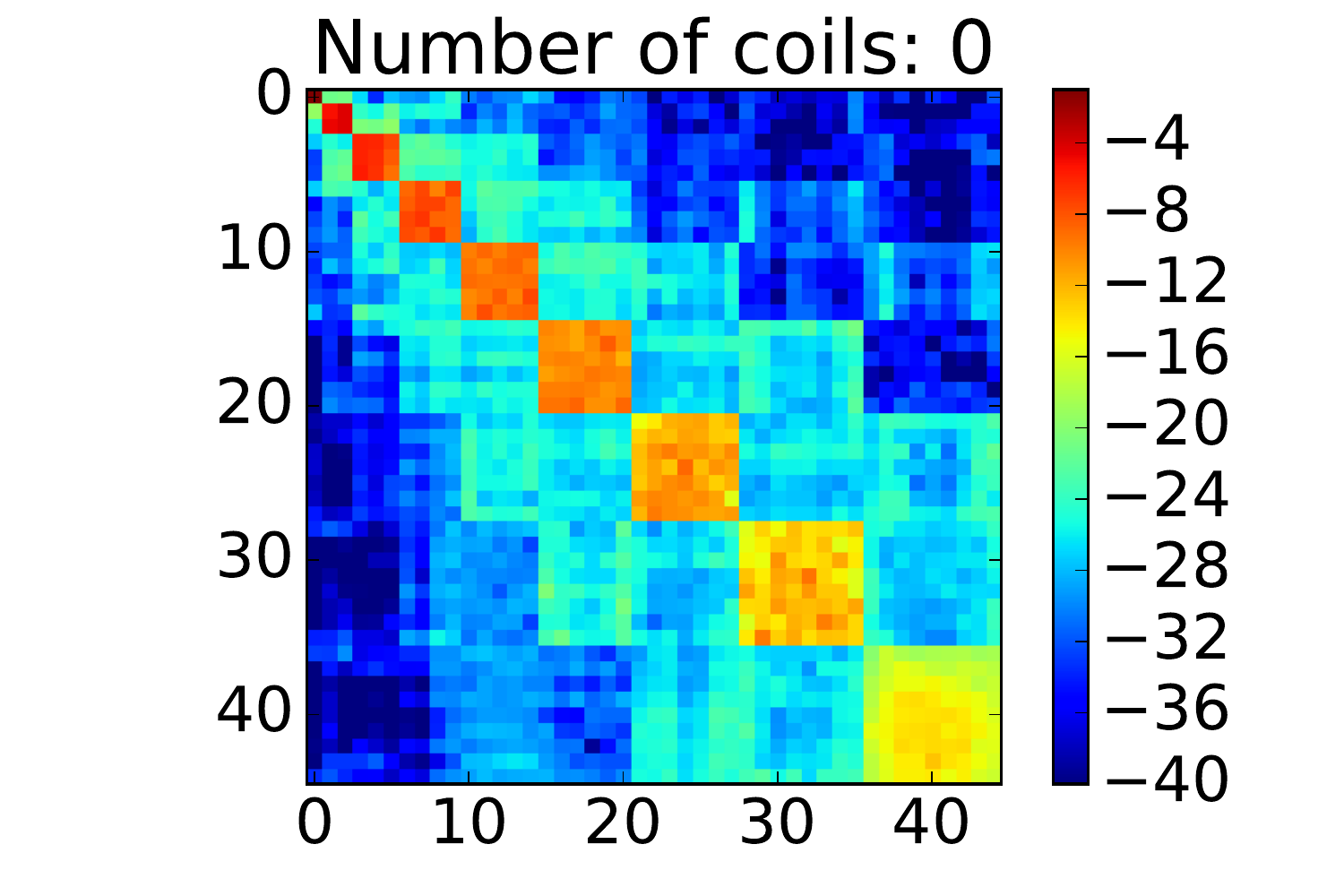}
		\caption{}\
	\end{subfigure}
	\begin{subfigure}[]{0.24\textwidth}
		\centering
		\includegraphics[trim = 1.7cm 0.5cm 1.2cm 0cm, clip, width=\textwidth]{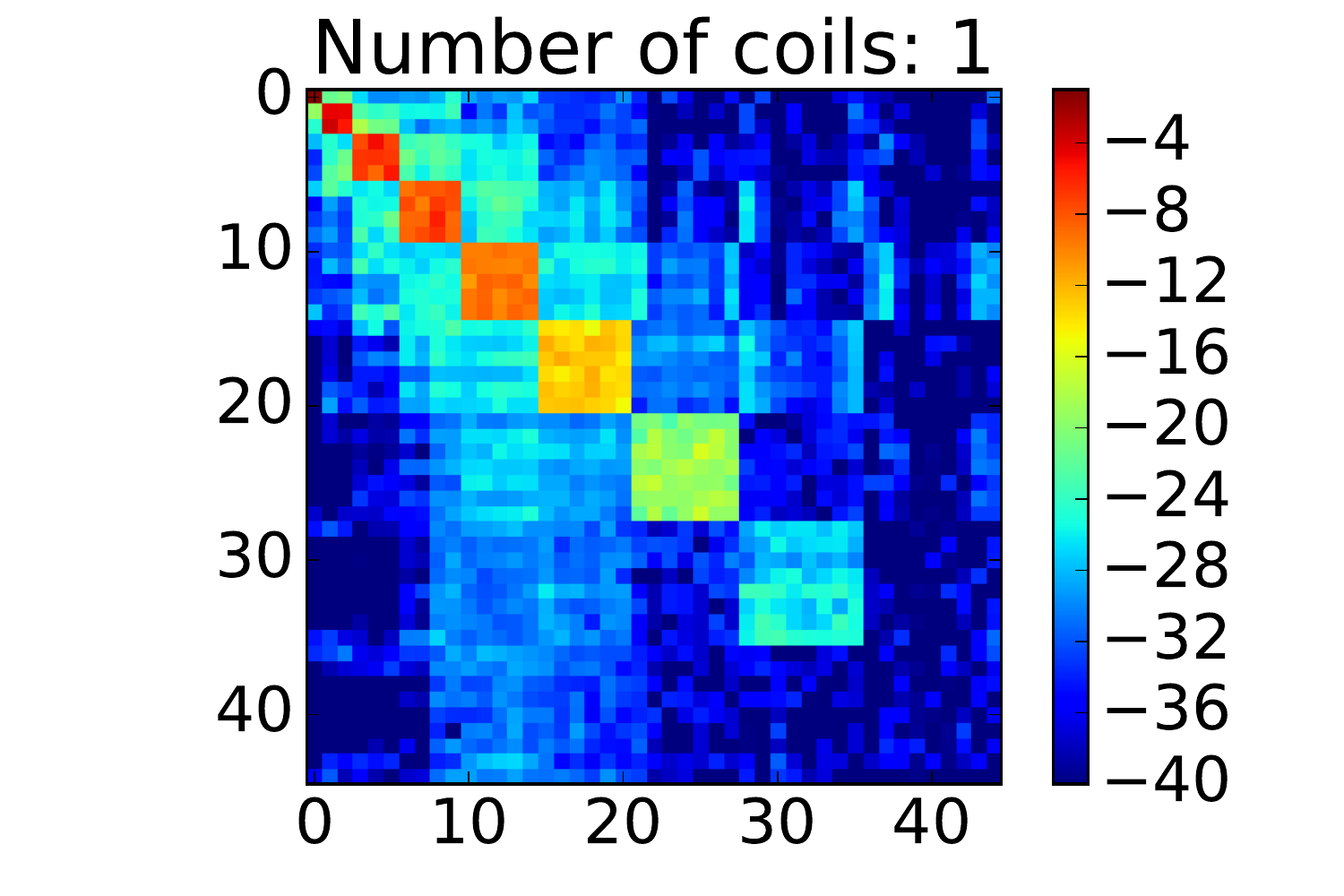}
		\caption{}\
	\end{subfigure}\\
	\begin{subfigure}[]{0.24\textwidth}
		\centering
		\includegraphics[trim = 1.7cm 0.5cm 1.2cm 0cm, clip, width=\textwidth]{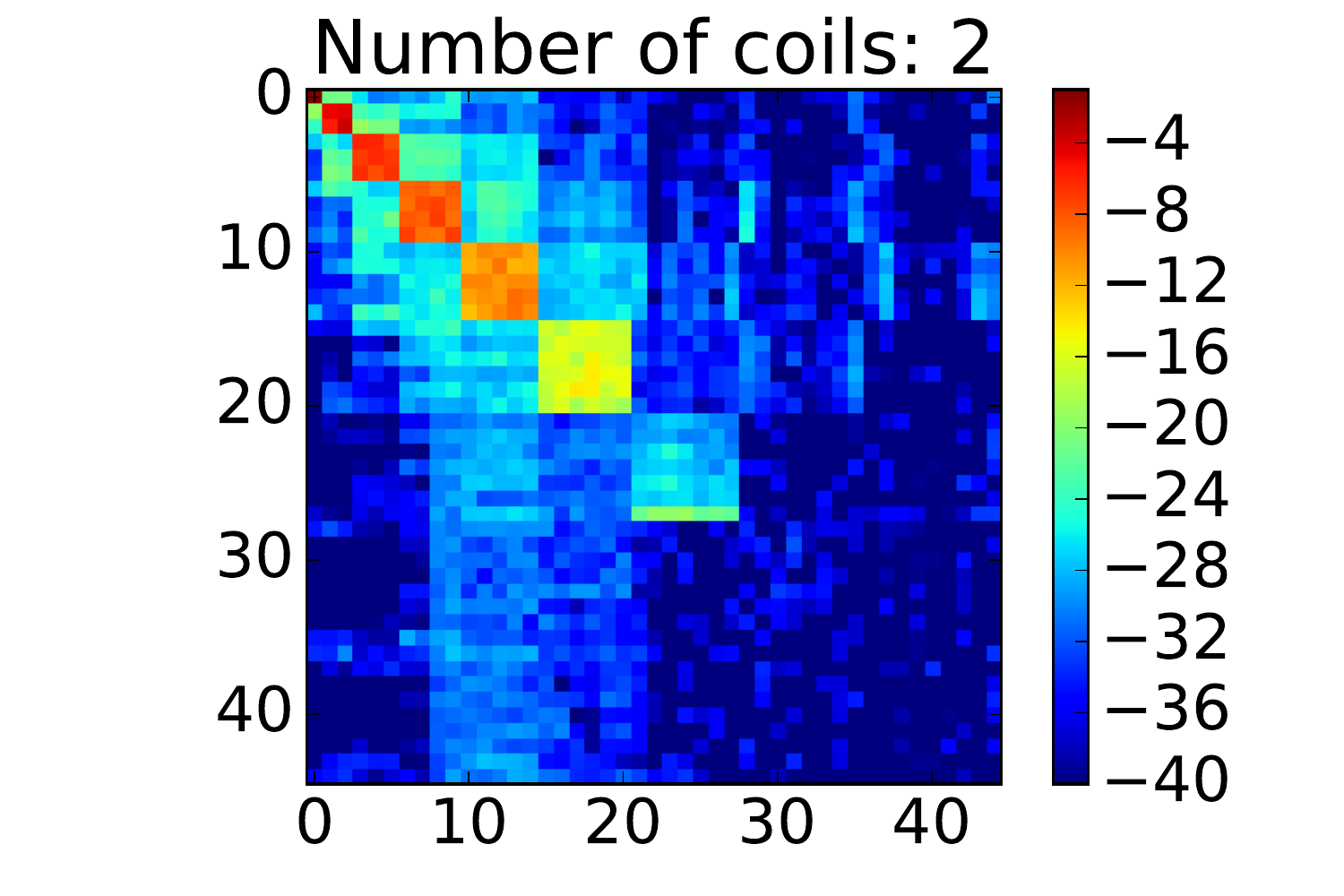}
		\caption{}\
	\end{subfigure}
	\begin{subfigure}[]{0.24\textwidth}
		\centering
		\includegraphics[trim = 1.7cm 0.5cm 1.2cm 0cm, clip, width=\textwidth]{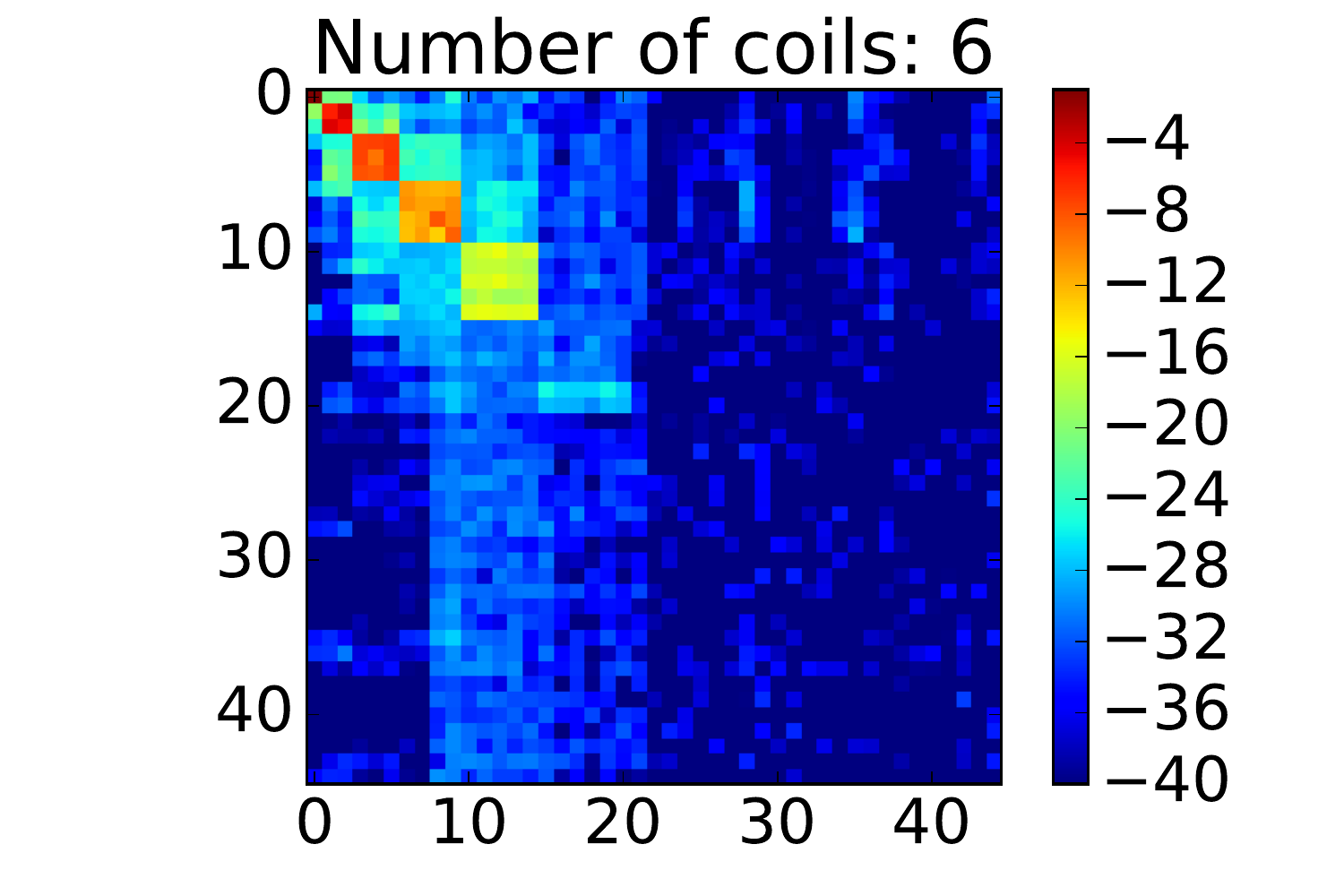}
		\caption{}\
	\end{subfigure}
	\caption{\label{fig:coilM6} Output power matrices in dBm of the 45-modes propagating in an OM2 fiber coiled around a 5~mm diameter rod for (a) 0, (b) 1 (c) 2 and (d) 6 numbers of coil.}
\end{figure}

Next, we experiment the same protocol with a 5~mm diameter rod. The significantly smaller bending radius induces a stronger impact on the modes' propagation. With only one coil, the propagation losses are sufficient to consider that the higher 9th order mode group is suppressed and the following 8th mode group reaches the noise floor with only 2 coils. We increase the number of coil up to 6 to visualize the influence on the lower order mode groups. We can observe that, even with a large number of coils, the bending has a negligible impact on the lowest 4 mode group orders. This means that the propagation of the first 10 HG modes is well guided, with a low inter-mode group XT even with very important bending. The transmission of the 5th order mode group is also maintained but with more losses. This demonstrate the possibility to confine and propagate the lowest 4 mode groups even with strong fiber bending, validating that 4 mode group MGDM as demonstrated in \cite{Lengle2016} can be implemented in a realistic fiber deployment scheme.\

%

\section{Conclusion}

To conclude, we have built a pair of 45-mode multiplexer and demultiplexer based on Multi-Plane Light Conversion addressing all the modes of a 50~$\mu$m core graded-index fiber. The multiplexers show an average insertion loss of 4~dB and an average cross-talk of -28~dB across the C-band. The performance, similar to previous 10 or 15-modes MPLC multiplexer, is enabled by using separable mode bases as the target output modes, which significantly reduces the required number of phase profiles. This selective 45-mode multiplexer was then used to explore the propagation properties of a standard OM2 fiber. We highlighted that the propagation-induced crosstalk between mode groups in a MMF is predominant with the nearest mode group. We finally studied the impact of bending on mode group loss and cross-talk, showing that mode groups up to the fourth order are preserved even with important bending. These two examples perfectly illustrate the new possibilities offered by this platform to investigate the modal properties of fiber and support the development of spatially multiplexed networks.

\end{document}